\documentclass[a4paper,12pt]{article}
\usepackage{latexsym,amssymb,amsfonts,amsmath,amsthm}
\usepackage[dvips]{graphicx}
\usepackage{comment}

\newtheorem{theorem}{Theorem}
\newtheorem{lemma}{Lemma}

\newtheorem{definition}{Definition}

\numberwithin{theorem}{section}
\numberwithin{lemma}{section}
\numberwithin{corollary}{section}
\numberwithin{proposition}{section}
\numberwithin{remark}{section}

\newcommand{\bs}[1]{\boldsymbol{#1}}

\usepackage{color}

\title{Comfortable place for quantum walkers on finite path}
\author{{\small 
Yoshihiro Anahara,$^{1}$ \quad
Norio Konno,$^{2}$\quad 
Hisashi Morioka,$^{3}$\quad
Etsuo Segawa$^{4}$
}\\ 
{\scriptsize $^{1}$ 
Collage of Engineering of Science, Yokohama National University, }\\
{\scriptsize Hodogaya, Yokohama, 240-8501, Japan.} \\
{\scriptsize $^{2}$ 
Department of Applied Mathematics, Faculty of Engineering
Yokohama National University }\\  
{\scriptsize Hodogaya, Yokohama, 240-8501, Japan.} \\
{\scriptsize $^{3}$ 
Graduate School of Science and Engineering, Ehime University, }\\  
{\scriptsize Bunkyo-cho 3, Matsuyama, Ehime, 790-8577, Japan.} \\
{\scriptsize $^{4}$ 
Graduate School of Environment and Information Sciences, Yokohama National University
}\\
{\scriptsize 
Hodogaya, Yokohama, 240-8501, Japan
} 
}
\date{}

\begin{document}
\maketitle

\begin{small}
\par\noindent
{\bf Abstract}. 
We consider the stationary state of a quantum walk on the finite path, where the sink and source are set at the left and right boundaries. 
The quantum coin is uniformly placed at every vertex of the path graph. At every time step, a new quantum walker penetrates into the internal from the left boundary and also some existing quantum walkers in the internal goes out to the sinks located  in the left and right boundaries. 
The square modulus of the stationary state at each vertex is regarded as the comfortability for a quantum walker to this vertex in this paper. We show the weak convergence theorem for the scaled limit distribution of the comfortability in the limit of the length of the path. 
\end{small}

\section{Introduction}
A primitive form of the dynamics of a quantum walk has appeared in \cite{FH}. A walker in this model is reflected and transmitted at each vertex of the one-dimensional lattice with some complex valued weight. In other word, each vertex plays a role of the locally scattering of a walker, where each vertex receives an inflow and also sends an outflow at each time step. It may be possible to say that such a time evolution of the whole system is extended to a unitary operator as the quantum walk in \cite{Gudder,Meyer}. The connection of quantum walks to a quantum graphs~\cite{Exner,Kuchment,GS}, which are  stationary Schr{\"o}dinger equation of the plain wave on a metric graph, can be seen in  \cite{Tanner,KHiguchi}, for example. 

The concept of quantum walks on a graph with global in- and out- flows toward the graph is proposed by \cite{FelHil1}. It is shown that this dynamics restricted to the internal graph converges to a stationary state~\cite{FelHil1,FelHil2,HS}. The scattering on the surface of the internal graph gives sometimes a structure of the internal graph~\cite{FelHil2,HSS3}. If we want to investigate the geometric information on the graph in more detail through this quantum walk model, it is natural to see the stationary state in the interior. The {\it comfortability} for quantum walker describes how many quantum walkers can stay in the interior
\cite{HSS3}. The comfortability extracts some interesting factors of graphs, for example, the complexity, odd-unicylic factor~\cite{HSS3}. 

If we fix the internal graph, the object of the comfortability for quantum walker is changed to the frequency of the inflow.
The situation that inserting the oscillated inflow into the internal graph from the outside can be translated into the situation ``rocking" the internal graph with this frequency~\cite{HS}. The frequency is described by $\xi\in \mathbb{R}/{2\pi \mathbb{Z}}$.  
In \cite{HKKMS}, the comfortability to the frequency for quantum walker on the finite path is computed: it is shown that the quantity of the comfortability can be tuned between $O(1)$ and $O(M^3)$, where $M$ is the length of the path, by adjusting the frequency $\xi$.
In this paper, we consider more detailed information of the comfortbility on the path; that is, 
\begin{center}
{\it Where is the comfortable place for quantum walkers on the path graph?} 
\end{center}
The square modulus of the stationary state at each place can be interpreted as the relative probability. 
Thus, we normalize the relative probability by the comfortability and regard it as the distribution of the comfortability. In this paper, we give an answer in the setting of the uniformly placed quantum coins under the concept of the distribution of the comfortability. We obtain the weak limit theorems of the distribution of the comforatability for large size $M$ which depends on the frequency.  
We find that the density function is expressed by a quite simple form depending on the frequency in this paper (see (\ref{eq:explicit}) for the explicit expression). 

We can notice that once a quantum walker penetrates into the interior, the situation for this quantum walker is the same as that of the absorption problem of quantum walks~\cite{ABNV}. The detailed analysis on the absorption problem especially the spectral analysis can be seen, for example, \cite{Parker}. Then, for a convergence time to the stationary state, such a spectral analysis on the truncated finite matrix of the time evolution operator of our model, which is non-unitary, will be useful.
We remain this problem in the interesting future's work. 

This paper is organized as follows. In Section~2, the definitions of this quantum walk model and the distribution of the comfortability.  
In Section~3, we provide the main result on the weak limit of the distribution of the comfortability for large size $M$. In Section~4, the proof of the main theorem is devoted. 

\section{Setting of our model}
The model consideded here is defined as follows. 
Let $C(j)$ be a two dimensional unitary matrix assigned at each vertex $j\in \mathbb{Z}$. Putting $|L\rangle=[1,0]^\top$ and $|R\rangle=[0,1]^\top$, we define \[P(j)=|L\rangle\langle L|C(j) \text{ and } Q(j)=|R\rangle\langle R|C(j).\]
The total state space treated here is the set of uniformly bounded functions such that   \[\ell^\infty(\mathbb{Z};\mathbb{C}^2)=\{\psi:\mathbb{Z}\to \mathbb{C}^2 \;:\; ||\psi(j)||_{\mathbb{C}^2}<c  \text{ for any $j\in\mathbb{Z}$}\}.\] Here $c>0$ is a finite constant depending on $j\in\mathbb{Z}$. The time evolution is the iteration of the unitary  operator $U_M$. More precisely, let $\psi_{t}\in \ell^\infty(\mathbb{Z};\mathbb{C}^2)$ be the $t$-th iteration of the quantum walk; then $\psi_{t+1}=U\psi_t$. 
Here $U_M$ is defined as follows: 
\[ (U_M\psi)(j) = P(j+1)\psi(j+1)+Q(j-1)\psi(j-1) \]
for any $j\in \mathbb{Z}$. 
Throughout this paper, we set the local unitary matrix $C(j)$ by 
\[ C(j)=\begin{cases}  
C_0=\begin{bmatrix} a & b \\ c & d \end{bmatrix} & \text{: $j\in\{0,\dots,M-1\}$,} \\
\\
I & \text{: otherwise.}
\end{cases} \]
We assume $abcd\neq 0$ to avoid the trivial walks. 
Note that the walk is free in the outside of the perturbed region $\{0,\dots,M-1\}$, that is, 
\begin{align*}
    (U_M\psi)(j)_L &= \psi(j-1)_L,  \text{ ($j\leq -2$ or $j\geq M-1$),} \\
    (U_M\psi)(j)_R &= \psi(j+1)_R,  \text{ ($j\leq 0$ or $j\geq M+1$).}
\end{align*} 
Here $\phi_J=\langle J|\phi\rangle$ for any $\phi\in\mathbb{C}^2$ ($J\in\{L,R\}$).  The initial state $\psi_0$ is set so that the perturbed region $\{0,\dots,M-1\}$ receives the inflow from the negative side with a frequency $\xi\in \mathbb{R}$ at every time step;
\[ \psi_0(j)=\begin{cases}
e^{i\xi j}|R\rangle & \text{: $j\leq 0$,}\\
0 & \text{: otherwise.}
\end{cases} \]
Once a quantum walker in the internal goes out to the outside, then it never goes back to the internal; such a quantum walker can be regarded as the outflow. 
Let $\phi_t:= e^{i\xi t}\psi_t$. Then $\phi_t$ converges to a stationary state at each vertex  
$j\in\mathbb{Z}$~\cite{FelHil1,FelHil2,HS}. 
Then in this paper, we are interested in the following normalized measure on the perturbed region $\{0,1,\dots,M-1\}$ which is the distribution of the comfortability. 
\begin{definition}
For any $j\in\{0,1,\dots,M-1\}$, we set the distribution of the comfortability on $\{0,1,\dots,M-1\}$ as follows: 
\[ \mu_M(j):=\lim_{t\to\infty}\frac{||\psi_{t}(j)||^2}{\sum_{j=0}^{M-1} ||\psi_{t}(j)||^2 }. \]
\end{definition}
\section{Main result}
Let the cumulative distribution of $\mu_M(\cdot)$ until $M x$ ($x\in \mathbb{R}$) be denoted by 
\[ F_M(x):=\sum_{j/M\leq x} \mu_M(j). \] 
Note that if $x<0$, then $F_M(x)=0$ and if $x>M-1$, then $F_M(x)=1$. We are interested in the shape of the derivative of $F_M(x)$ on the normalized region $x\in[-1,1]$ for large $M$ because the shape tells us the comfortable place for the quantum walker. If we put $X_M$ as a random variable following the distribution $\mu_M$, then the scaled random variable $X_M/M$ follows the distribution $F_M(x)$. Our interest is the limit of $F_M(x)$ as $M\to\infty$. In particular, if $F_M(x)$ converges to $F_*(x)$ in the limit of large $M$ for each $x\in[0,1]$, then the scaled random variable $X_M/M$ converges to $Z$ in distribution, where $Z$ follows 
a limit distribution $F_*$; in this paper, we describe it by 
\[ X_M/M \stackrel{d}{\to} f(x)\;\; (M\to\infty)\]
with the density function $f(x)$  of the limit distribution $F_*(x)$. 

We use the parameter $\omega=\arg(\det(C_0))/2+\xi$ instead of the frequency of the initial state $\xi$. The unit circle in the complex plain is divided into the following three parts: 
\[ B_{out}=\{\omega :\; |\cos \omega|>|a|\};\; \partial B=\{\omega :\; |\cos \omega|=|a|\};\; B_{in}=\{\omega ,\; |\cos \omega|<|a|\}.\]
Moreover we newly introduce the parameter $\theta$ in the case of $\omega\in B_{in}$: 
\begin{equation} \label{eq:theta}
\theta=
\begin{cases}
\arccos \frac{\cos \omega}{|a|} & \text{: $ \frac{\cos \omega}{|a|}>0$,}\\
\pi-\arccos \frac{\cos \omega}{|a|} & \text{: otherwise.}
\end{cases} 
\end{equation}
The absolute value $|\theta|$ represents how the input parameter $\omega\in B_{in}$ is closed to the boundary $\partial B$. 
In this paper, the quantity of the parameter $\theta$ changes depending on the size $M$ satisfying $\lim_{M\to \infty}M|\theta|=\theta_*$ with some value $\theta_*\in[0,\infty]$. The dependence of $\theta$ is divided into the following three situations: 
\begin{center}
(i) $\theta_*=0$; (ii) $0<\theta_*<\infty$; (iii)  $\theta_*=\infty$. 
\end{center}
Case (i) realizes the case that the parameter $\theta$ is located in quite close place to $\delta B$.  
Case (ii) realizes also the case that the parameter $\theta$ is closed to $\delta B$, but its distance is $O(1/M)$.  
Case (iii) realizes the case that the parameter $\theta$ is placed around the middle of $B_{in}$.  
Then we describe  the cases (i), (ii) and (iii) by 
$|\theta|\ll 1/M$, $|\theta|\asymp 1/M$ and $|\theta|\gg 1/M$, respectively. 
Now we are ready to state our  main theorems. 
\begin{theorem}\label{thm:main}
Let $X_M$ be the random variable following the stationary distribution with the size $M$ and with the input parameter $\omega$. 
Let $\bs{1}_{[0,1]}(x)$ be the indicator function between $0$ to $1$. 
Then we have 
\[ X_M/M \stackrel{d}{\to} \rho^{(\omega)}(x), \;(M\to\infty) \]
where $\rho^{(\omega)}(x)$ is described up to the input parameter $\omega$ as follows. 
\begin{equation}\label{eq:explicit}
    \rho^{(\omega)}(x)=\bs{1}_{[0,1]}(x)\times 
    \begin{cases}
    \delta(x) & \text{: $\omega\in B_{out}$,}\\
    3(1-x)^2 & \text{: ``$\omega\in \partial B$" or ``$\omega\in B_{in }$ and $|\theta|\ll 1/M$",} \\
    c(\theta_*) \sin^2[(1-x)\theta_*] & \text{: $\omega\in B_{in}$ and $|\theta|\asymp 1/M$,}\\
    1 & \text{: $\omega\in B_{in}$ and $|\theta|\gg 1/M$,}
    \end{cases}
\end{equation}
where $c(\theta_*)$ is the normalized constant; that is, 
\[c(\theta_*)=2\left(1-\frac{\sin 2\theta_*}{2\theta_*}\right)^{-1}.\] 
\end{theorem}

\noindent
If $|\theta|\asymp 1/M$, then Theorem~\ref{thm:main} implies that the cumulative limit distribution can be computed by   
\begin{align}
F_{\theta_*}(y) 
&:= \int_{0}^y \rho^{(\omega)}(x)dx \notag\\
& = \frac{1}{1-\frac{\sin 2\theta_*}{2\theta_*}}\left( y+ \frac{\sin(2\theta_*(1-y))}{2\theta_*} -\frac{\sin 2\theta_*}{2\theta_*}\right). \label{eq:distlibution}
\end{align}
The limit $\theta_*\to 0$ corresponds to the case for $|\theta|\ll 1/M$ while the limit $\theta_*\to \infty$ corresponds to the case for $|\theta_*|\gg 1/M$. Indeed, by  (\ref{eq:distlibution}), we obtain
\begin{align}
    \lim_{\theta_*\to 0}F_{\theta_*}(y)&=y^3-3y^2+3y, \label{eq:poly}\\
    \lim_{\theta_*\to \infty}F_{\theta_*}(y)&=y, \label{eq:uniform}
\end{align}
which are nothing but the cumulative limit distribution for the cases of 
$|\theta|\ll 1/M$ and $|\theta|\gg 1/M$ in Theorem~\ref{thm:main}. This shows the continuity of $F_{\theta_*}(y)$ with respect to  $\theta_*\in[0,\infty]$. 

In the case of $\omega\in B_{out}$, there should be the optimal normalization order which is greater than $O(1/M)$ because the limit density is the delta function in the $1/M$ normalization. Indeed, $X_M$ converges without any normalization as follows. 
\begin{theorem}\label{thm:Bout}
If the input is $\omega \in B_{out}$, then 
\[\lim_{M\to\infty}\mu_M(j)=(1-\lambda_+^{-2})\lambda_+^{-2j}\]
for any $j\in\{0,1,2,\dots\}$.
Here $\lambda_+$ is the solution of the following quadratic equation with $|\lambda_+|>1$:
\[ \lambda^2 -2\frac{\cos \omega}{|a|} \lambda + 1=0. \]
\end{theorem}
\section{Proof of theorems}
First we prepare the following known result. 
\begin{lemma}[\cite{HKKMS}]\label{lem:stationarystate}
Let $\phi$ be the stationary state. Then the relative probability at $n\in\{0,1,\dots,M-1\}$ is described by  
\begin{equation}\label{eq:stationary1}
    ||\phi(n)||^2=
          \frac{1}{|a|^2+|b|^2\zeta^2(M)}\left(\;|a|^2+|b|^2\zeta^2(M-n-1)+|b|^2\zeta^2(M-n)\;\right), 
\end{equation}
where 
\[\zeta(m)=\mathcal{U}_{m-1}\left(\frac{\omega+\omega^{-1}}{2|a|}\right), \]
and $\mathcal{U}_m(\cdot)$ ($m=0,1,2,\dots$) is the Chebyshev polynomial of the second kind. 
\end{lemma}
From this Lemma, using the properties of the Chebyshev polynomial, we can compute the comfortability as follows. 
\begin{lemma}[\cite{HKKMS}]\label{lem:comf}
The comfortability of quantum walker is defined as follows. 
\[ \mathcal{E}_M(\omega):=\sum^{M-1}_{n=0} ||\phi(n)||^2.\]
Then we have 
\begin{equation}\label{eq:energy}
    \mathcal{E}_M(\omega)=
    \frac{1}{|a|^2+|b|^2\zeta^2(M)}\left\{
    M|a|^2+\dfrac{|b|^2}{\left(\lambda_{+}-\lambda_{-}\right)^2}\left(\zeta^2(M+1)-\zeta^2(M-1)-4M\right)\right\},
\end{equation}
where $\lambda_\pm$ is the solution of the following quadratic equation with $|\lambda_+|>1$ and $|\lambda_-|<1$:
\[ \lambda^2 -2\frac{\cos \omega}{|a|} \lambda + 1=0. \]
Here the comfortability $\mathcal{E}_M(\omega)$ is continuous at $\omega_*\in \partial B$; that is,  
\[\mathcal{E}_M(\omega_{*})=\lim_{\omega\to\omega_*}\mathcal{E}(\omega)=\dfrac{1}{3}\dfrac{M}{|a|^2+|b|^2 M^2}\left(3|a|^2+|b|^2+2|b|^2 M^2\right).
\]
\end{lemma}

\noindent Now let us see the  proof of Theorems~\ref{thm:main} and \ref{thm:Bout}.
\begin{proof}
Since the stationary distribution is 
\[ \mu_M(n)=\frac{||\phi(n)||^2}{\mathcal{E}_M(\omega)}, \]
we insert (\ref{eq:stationary1}) and (\ref{eq:energy}) into this directly and compute the cumulative distribution $F_M(x)$, we obtain the conclusions. Let us see it case by case. \\

\noindent (1) $\omega\in B_{out}$ case (proofs of Theorem~\ref{thm:main} for $\omega\in B_{out}$ and Theorem~\ref{thm:Bout}): \\
It is enough to show Theorem~\ref{thm:Bout}. 
Let us consider the case for $\omega\in B_{out}$. 
We have 
\[ \zeta_{M-m}=\frac{\lambda_+^{M-m}-\lambda_-^{M-m}}{\lambda_+-\lambda_-}\sim \frac{\lambda_+^{M-m}}{\lambda_+-\lambda_-} \]
because $|\lambda_+|>1>|\lambda_-|$ when $\omega\in B_{out}$. 
Using this, we have
\begin{align*}
    ||\phi(n)||^2 &\sim \frac{1}{|a|^2+|b|^2\left(\frac{\lambda_+^{M}}{\lambda_+-\lambda_-} \right)^2}\left\{ |a|^2+|b|^2\left( \left(\frac{\lambda_+^{M-n-1}}{\lambda_+-\lambda_-}\right)^2+\left(\frac{\lambda_+^{M-n}}{\lambda_+-\lambda_-}\right)^2 \right) \right\}\\
    &\sim (1+\lambda_+^2)\lambda_+^{-2(n+1)}
\end{align*}
because $|b|\neq 0$. 
Thus by the normalization, we obtain the desired conclusion of Theorem~\ref{thm:Bout}. 
\\

\noindent (2) $\omega\in B_{in}$ case (proof of Theorem~\ref{thm:main}):\\
The computational method are quite similar for each case; $|\theta|\ll 1/M$, $|\theta|\asymp 1/M$, $|\theta|\gg 1/M$. 
So let us show only the $\omega\in B_{in}$, $|\theta|\asymp 1/M$ case. The consistency of the other cases for $|\theta|\ll 1/M$ and  ``$|\theta|\gg 1/M$ or $\omega\in \delta B$" may be able to be confirmed by (\ref{eq:poly}) and (\ref{eq:uniform}), respectively. 

We concentrate on the difference between $\rho^{(\omega)}$ and the distribution $\mu_M$ as follows:  
\[ \Delta_M(n):=\left|\mu_M(n)-\frac{1}{M}\rho\left(\frac{n}{M}\right)\right|. \]
First let us show $\{M\Delta_M(n)\}_{M}$ uniformly converges to $0$ in the limit of $M\to\infty$; that is,   \begin{equation}\lim_{M\to \infty}M\max_{0\leq n\leq M}(\Delta_M(n))\to 0. \end{equation}
By Lemmas~\ref{lem:stationarystate} and \ref{lem:comf}, the distribution $\mu_M(n)$ can be expressed by $\mu_M(n)=B_M(n)/A_M$, where 
\begin{align*}
    A_M &= (|b|^2+|a|^2 \sin^2\theta)M-\frac{|b|^2}{4}\frac{\sin 2 M\theta \sin 2\theta}{\sin^2\theta}, \\
    B_M(n) &= |a|^2\sin^2 \theta +|b|^2 \sin^2 (M-n-1)\theta + |b|^2\sin^2 (M-n)\theta. 
\end{align*}
In the following, let us obtain the lower and upper bounds of $A_M$ and $B_M(n)$, respectively. 
{\bf Estimation of $A_M$}: 
We put $\theta_*=M\theta$. 
Then 
\[ \frac{\sin 2M\theta \sin 2\theta}{\sin^2\theta}=2\sin 2\theta_* \frac{\cos\theta}{\sin\theta}. \]
Using the inequality 
\[ \theta-\theta^3/6 < \sin \theta < \theta\text{ and }1-\theta^2/2 < \cos \theta <1, \;(0<\theta<\pi)\]
we see 
\[ 2\sin 2\theta_* \left( \frac{1}{\theta}-\frac{1}{2}\theta \right)<\frac{\sin 2M\theta \sin 2\theta}{\sin^2\theta}<2\sin 2\theta_* \left( \frac{1}{\theta}+\frac{1}{5}\theta \right). \]
By inserting $\theta=\theta_*/M$, 
there exist constant values $c_\pm$ such that 
\begin{equation}\label{eq:delta1}
  c_-/M<  \Delta^{(1)}_M<c_+/M,
\end{equation}
where $\Delta^{(1)}_M$ is the difference between $A_M$ and the denominator of $(1/M)\;\rho(n/M)$; that is,  
\[\Delta^{(1)}_M:=A_M-M|b|^2\left(1-\frac{\sin 2\theta_*}{2\theta_*}\right).\] 
\\
{\bf Estimation of $B_M(n)$}: 
There exist $c_1$ and $c_2$ such that $c_1 \delta <\sin^2(x-\delta)-\sin^2x< c_2\delta$, we have 
\begin{align*}
    B_M(x) &\leq |a|^2\theta^2+|b|^2\sin^2[(1-n/M)\theta_*-\theta]+|b|^2\sin^2[(1-n/M)\theta_*] \\
    &\leq |a|^2\theta^2+|b|^2\{\sin^2[(1-n/M)\theta_*]+c_2\theta \}+|b|^2\sin^2[(1-n/M)\theta_*] \\
    &\leq 2|b|^2 \sin^2[(1-n/M)\theta_*]+c_2'\theta,
\end{align*}
where $c_2'$ is a constant value, for example, we can put $|b|^2c_2+|a|^2$. 
On the other hand, 
\begin{align*}
    B_M(x) &\geq |a|^2\left( \theta-\frac{\theta^3}{6} \right)+|b|^2 \left\{ \sin^2[(1-n/M)\theta_*+c_1\theta] \right\}+|b|^2\sin^2[(1-n/M)\theta_*] \\
    &\geq 2|b|^2\sin^2[(1-n/M)\theta_*]+c_1'\theta, 
\end{align*}
where $c_1'$ is a constant value, for example, $|b|^2 c_1$. 
Inserting $\theta=\theta_*/M$, we obtain
\begin{equation}\label{eq:delta2}
 c'_-/M<\Delta^{(2)}_M(n)<c'_+/M,
\end{equation}
where $\Delta^{(2)}(n)$ is the difference between $B_M(n)$ and the numerator of $(1/M)\rho(n/M)$; that is, 
\[\Delta^{(2)}_M(n):=B_M(n)-2|b|^2\sin^2[(1-n/M)\theta_*].\]
Here $c_{\pm}'$ are constant value which is independent of $n$ and $M$. \\

\noindent From (\ref{eq:delta1}) and (\ref{eq:delta2}), we obtain 
\begin{align*}
    \frac{B_M(n)}{A_M} &= \frac{2|b|^2\sin^2[(1-n/M)\theta_*]+\Delta^{(2)}_M(n)}{M|b|^2(1-\sin (2\theta_*)/(2\theta_*))+\Delta_M^{(1)}} \\
    &= \frac{1}{M}\rho_M(n/M)+O\left(\frac{1}{|b|^2(1-\sin 2\theta_*/(2\theta_*))}\frac{\Delta_M^{(2)}(n)}{|M|}\right).
\end{align*}
Note that the second term in RHS is nothing but $\Delta_M(n)$ and can be uniformly bounded by $c'/M^2$ with some constant value $c'$ which is independent of $n$ and $M$ by (\ref{eq:delta2}). 
Then we obtain
\[ \lim_{M\to\infty}M\max_{0\leq n\leq M}\Delta_M(n)= 0. \]
Therefore
\begin{align*}
    \left|\sum_{n\leq Mx}\mu_M(n)-\sum_{n\leq Mx}\frac{1}{M}\rho^{(\omega)}(n/M)\right| &< \sum_{n\leq Mx}\Delta_M(n)<\left(\max_{0\leq n\leq M}\Delta_M(n)\right)\cdot Mx\stackrel{(M\to\infty)}{\longrightarrow} 0.
\end{align*}
Then we have for any $x\in \mathbb{R}$, 
\[ \sum_{n\leq Mx}\mu_M(n)\to \int_{-\infty}^x \rho^{(\omega)}(s)ds\;\;(M\to\infty). \]
\end{proof}
\section{Summary and discussion}
In this paper, we considered the weak limit theorems for the distribution of the comfortability on the path graph with respect to the length of the path. 
Assume $\det(C_0)=1$ for a simplicity. 
The limit distribution depends on the frequency of the rocking toward the graph. If the frequency $\omega\in B_{in}$, we used the deformed parameter $\theta$ defined by 
\[ \theta=\begin{cases} \theta' & \text{: $\cos\theta'>0$, }\\ \pi-\theta' & \text{: otherwise.}\end{cases}\]
Here \[\cos\theta'=\frac{\cos \omega}{|a|},\;\sin\theta'\geq 0.\]
Since the limit distribution can be described without any dependency on the parameters of the quantum coin, the limit distribution reflects a universal property of the quantum walk on the one-dimensional lattice.

Finally, let us discuss on the comparison between our results and the well-known limit distribution of quantum walk on $\mathbb{Z}$ obtained by \cite{Konno}. 
Let $U'_M$ be defined by $(U'_M\phi)(j)=(U_M\phi)(j-[M/2])$. 
The limit distribution on $\mathbb{Z}$ can be defined by  
\[ G(x)=\lim_{t\to\infty}\left(\sum_{j/t< x} \lim_{M\to\infty}||\;({U_M'}{}^{t}\varphi_0)(j)\;||^2\right).\]
Here the initial state is set  $\varphi_0(x)=\delta_{[M/2]}(x)[\alpha, \beta]^\top$; this means that the whole space is denoted by an $\ell^2$-space and the inflow is the one-shot at time $0$ in the middle of the $\{0,\dots,M-1\}$. 
On the other hand, the limit distribution obtained by this paper was defined by 
\[F(x)=\lim_{M\to\infty}\frac{1}{{\mathcal{E}_M}(\omega)}\left(\sum_{0<j/M<x} \lim_{t\to\infty}||\;(U^t_M\psi_0)(j)\;||^2 \right). \]
Here the support of the initial state $\psi_0$ is outside of $\{0,1,\dots,M\}$ so that the internal receives the inflow at every time step with the frequency $\omega$  and the whole space is considered in an  $\ell^\infty$-space. 
Thus the order of the spatial and temporal limits are reversed each other.  
Let us find some connections between them. 
First, we can notice that the value $e^{i\omega}$ can be translated into the eigenvalue of the time evolution operator $U'$ in the Fourier space with the wave number $\theta'$. 
Thus $\omega$ corresponds to the pseudo-energy while $\theta'$ corresponds to the pseudo-momentum.  
Moreover the density function of $G(x)$, $dG(x)/dx$, is known as
\[ K(x)=\frac{\sqrt{1-|a|^2}}{\pi (1-x^2)\sqrt{|a|^2-x^2}}\bs{1}_{(-|a|,|a|)}(x) \]
by \cite{Konno}. 
Interestingly, if we set the symmetric condition~\cite{Konno}, ``$|\alpha|=|\beta|$ and $\mathrm{Re}(a\alpha \overline{b\beta})=0$", then this curve can be expressed by the following parametric expression:
\[ K:\;\left\{ \left( \frac{\partial\omega}{\partial\theta'},\;\frac{1}{\pi|\partial^2\omega/\partial\theta'{}^2|} \right)\;:\;\theta'\in[0,2\pi) \right\}. \]

On the other hand, putting $\psi_t:=U_M^t\psi_0$, 
we have $\exists \lim_{t\to\infty}e^{it\omega}\psi_t=:\psi_*$ in pointwise, and the function $\psi_*\in \ell^\infty(\mathbb{Z};\mathbb{C}^2)$ satisfies the following generalized eigenequation~\cite{FelHil2,HS}: 
\[ U_M\psi_*=e^{-i\omega}\psi_*. \]
Therefore the limit distribution on $\mathbb{Z}$ , where a quantum walker never feel the boundaries throughout the time evolution, derives from the generalized eigenvalue of the time evolution operator while the limit distribution treated in this paper, where a quantum walker feels the boundaries at every time step, derives from its generalized eigenvector. We expect that further consideration to such connections will reveal more fundamental structure of quantum walks in the future. \\

\noindent{\bf Acknowledgements.} H. M. was supported by the grant-in-aid for young scientists No.~16K17630, JSPS. 
E.S. acknowledges financial supports from the Grant-in-Aid of
Scientific Research (C) Japan Society for the Promotion of Science (Grant No.~19K03616) and Research Origin for Dressed Photon.\\


\end{document}